\documentclass[english]{article}
\usepackage[T1]{fontenc}
\usepackage[latin9]{inputenc}
\usepackage{amsmath}
\usepackage{graphicx}
\usepackage{bbold}
\usepackage{tikz}
\usepackage{lipsum}

\usepackage[hyphens]{url}
\PassOptionsToPackage{hidelinks}{hyperref}

\makeatletter

\newcommand{\nc}{\newcommand}
\newcommand{\rc}{\renewcommand}
\nc{\mc}{\mathcal}
\nc{\op}[1]{\operatorname{#1}}
\nc{\opcat}[1]{\mathbf{#1}}
\rc{\t}[1]{\text{#1}}

\nc{\id}{\op{id}}

\nc{\umutnote}[1]{{\marginpar{\small \textcolor{blue}{#1}}}}

\nc{\cA}{\mc{A}}\nc{\cB}{\mc{B}}\nc{\cC}{\mc{C}}\nc{\cD}{\mc{D}}\nc{\cE}{\mc{E}}\nc{\cF}{\mc{F}}\nc{\cG}{\mc{G}}\nc{\cH}{\mc{H}}\nc{\cI}{\mc{I}}\nc{\cJ}{\mc{J}}\nc{\cK}{\mc{K}}\nc{\cL}{\mc{L}}\nc{\cM}{\mc{M}}\nc{\cN}{\mc{N}}\nc{\cO}{\mc{O}}\nc{\cP}{\mc{P}}\nc{\cQ}{\mc{Q}}\nc{\cR}{\mc{R}}\nc{\cS}{\mc{S}}\nc{\cT}{\mc{T}}\nc{\cU}{\mc{U}}\nc{\cV}{\mc{V}}\nc{\cW}{\mc{W}}\nc{\cX}{\mc{X}}\nc{\cY}{\mc{Y}}\nc{\cZ}{\mc{Z}}

\nc{\PP}{\mathbb{P}}
\nc{\ZZ}{\mathbb{Z}}
\nc{\RR}{\mathbb{R}}
\nc{\bbC}{\mathbb{C}}
\nc{\CC}{\mathbb{C}}

\usepackage{INTERSPEECH2020}

\makeatother

\usepackage{babel}
\begin{document}
\address{}
\email{}
\name{Umut Isik, Ritwik Giri, Neerad Phansalkar, Jean-Marc Valin, \\ Karim Helwani, Arvindh Krishnaswamy}
\address{Amazon Web Services}
\email{umutisik@amazon.com, ritwikg@amazon.com}

\title{PoCoNet: Better Speech Enhancement with Frequency-Positional Embeddings, Semi-Supervised Conversational Data, and Biased Loss}

\maketitle
\begin{abstract}
    Neural network applications generally benefit from larger-sized models, but for current speech enhancement models, larger scale networks often suffer from decreased robustness to the variety of real-world use cases beyond what is encountered in training data. We introduce several innovations that lead to better large neural networks for speech enhancement. The novel PoCoNet architecture is a convolutional neural network that, with the use of frequency-positional embeddings, is able to more efficiently build frequency-dependent features in the early layers. 
    A semi-supervised method helps increase the amount of conversational training data by pre-enhancing noisy datasets, improving performance on real recordings. A new loss function biased towards preserving speech quality helps the optimization better match human perceptual opinions on speech quality. Ablation experiments and objective and human opinion metrics show the benefits of the proposed improvements.  

\end{abstract}

\section{Introduction}

Neural network based approaches have greatly improved the output quality of speech enhancement systems \cite{xia2013speech, xu2014regression, weninger2015speech, han2015learning, attentionwaveunet}. These networks are trained, typically, in a supervised setting, with synthetic mixtures of clean speech and known noise clips, sometimes with synthetic reverberation added. Usually, the model is used to estimate a magnitude gain for each bin in the time-frequency domain representation of the noisy and/or reverberant mixture signal. Recent \emph{phase-aware} models use a complex ratio mask instead of magnitude gain \cite{williamson2015complex, le2019phasebook}, while other approaches work directly in the waveform domain \cite{rethage2018wavenet, macartney2018improved, germain2018speech}. 

The speech enhancement problem has multiple challenges associated with it. First, the model needs to be robust to the multitude of different speech, recording, and noise conditions present in real-world usage. Second, clean speech data for training is limited in the public domain, with the biggest datasets coming from read material. Third, the task becomes increasingly difficult in low signal-to-noise ratio (SNR) cases, which can be helped by training larger models, which in turn makes the model more prone to fitting to the biases of the available dataset, decreasing robustness to other real-world conditions, making both of the first two challenges more pronounced. And fourth, the mismatch between human perception of sound quality and standard loss functions and metrics \cite{hu2007evaluation} can make well-optimized models perform worse in human evaluation. 

We propose several architectural, data preparation, augmentation and loss-function innovations that help meet the above stated challenges for large neural networks for speech enhancement. 

On convolutional architectures, standard implementations in the time-frequency domain rely on 1D or 2D convnets. In the typical 1D architecture (e.g. ConvTASNet \cite{luo2019conv}), the kernels move in the time-direction, and are fully connected in the frequency direction. These tend to have very large weight matrices in the early layers, where the architecture could benefit from a more hierarchical development of features. On the other hand, in standard 2D U-Net models where kernels move in both the time and frequency directions \cite{park2017fully}, early layer activations are blind to what frequency they operate in -- even in the case when padding is used, these early features' receptive fields have not yet reached the edges of the time-frequency image. Our proposed architecture has the advantages of both options, it is a 2D U-Net (with DenseNet blocks and self-attention) with small kernels, and can therefore develop features hierarchically, but can also take into account frequency information in early layers with the inclusion of frequency-positional embeddings. 

On the data front, we scale up the amount of clean conversational data available for training by using a semi-supervised approach. The clean portion of the LibriSpeech dataset, our starting point, contains data only from audio books, which is not conversational. The larger VoxCeleb dataset \cite{nagrani2017voxceleb}, on the other hand, is from television broadcasts, and contains background music and effects, some of the data is also highly reverberant. We use LibriSpeech-trained speech enhancement models to isolate the clean speech in VoxCeleb2 and eliminate reverberant clips, and show that adding this processed clean speech dataset to the training data improves the robustness of the model to conditions not well-represented in LibriSpeech. To make the most effective use of the data, we also use an extensive data augmentation stack that also helps address specific failure modes. 

We also apply synthetic reverberation in the dataset using a library of recorded and synthetically generated room impulse responses. We train separate models to target the task with and without partial dereverberation. 
For non-dereverberating models, reverberation is added during training to the clean speech data as an augmentation before mixing. For training partially dereverberating models, we add, to the clean speech labels, a faster decaying version of the reverberation as was done in \cite{zhao2018late}.

We use L1 losses across the board to help deal with dataset noise. We use a linear combination of two losses. The first is a new L1-loss on magnitudes which is biased to penalize under-estimation of speech time-frequency bin magnitudes, as well as weighted towards high-frequencies, which makes the output of the trained model better preserve speech quality and avoid muffling. The second, is an L1 loss in the audio waveform domain, which is backpropagated though the STFT layer and complex multiplication to the estimated complex ratio mask values in the time-frequency domain. 

To measure the performance of our model, we rely on Mean Opinion Score (MOS) subjective testing crowd-sourced on Amazon Mechanical Turk, using model outputs on the Deep Noise Suppression (DNS) challenge \cite{dnschallengefinal} pre-competition test set, as well as on standard numerical metrics on the synthetic portion of the same test set. An ablation study shows the added improvement to human MOS and numerical metrics from each proposed component discussed above. 

\section{Method}

Let $s$ be the clean speech audio signal and $x = s * h + n$ be the same signal with added noise $n$ and reverberated version $s * h$, which is convolved with a room impulse response $h$, and let $y$ be the denoised and/or dereverberated target signal. The neural model $\mathcal{N}$ takes as input the STFT of the reverberant and noisy example $s * h + n$ and estimates the complex ratio mask that would give the target signal estimate as: $$\widehat{y} = \op{ISTFT}(\mathcal{N}(\op{STFT}(x)) \cdot STFT(x)).$$

\begin{figure*}[h]

\begin{center}
\end{center}

\begin{tikzpicture}
     \node (fig1) at (0,0)
     {\includegraphics[width=1\textwidth,height=1.65in]{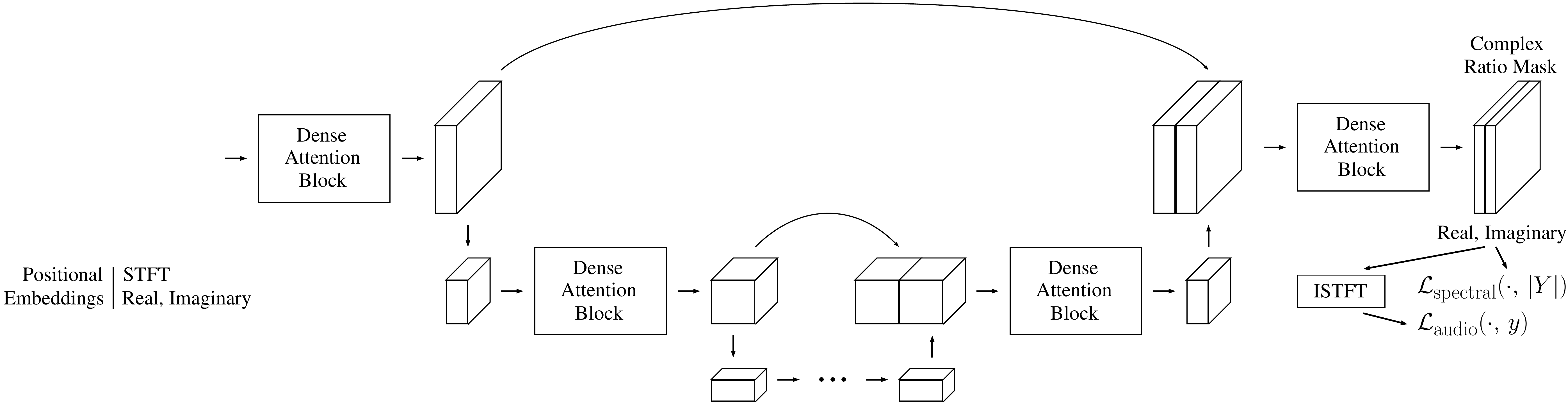}};

       \node[transform shape] at (-7.11,0.52) {\includegraphics[width=0.75in,height=0.906in]{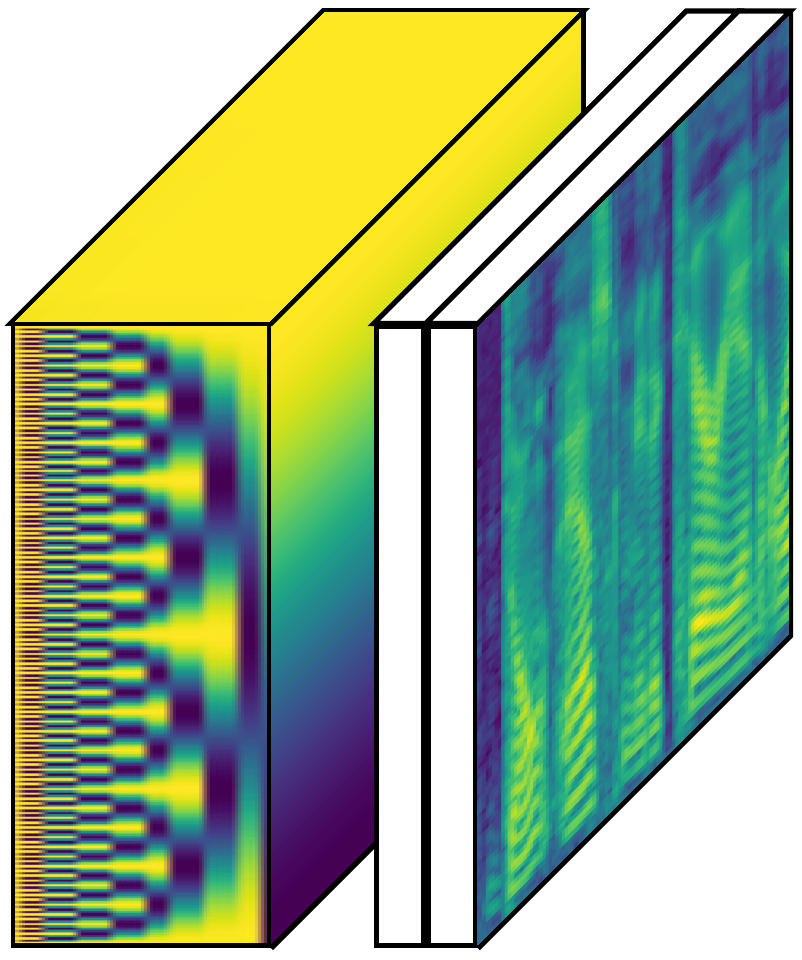}};
\end{tikzpicture}

\caption{Top two levels of the U-Net architecture shown with frequency-positional embeddings and STFT  real and imaginary parts inputs; and real and imaginary parts of complex ratio mask outputs. We use a 6-level U-Net architecture.}
\label{figunetandpos}
\end{figure*}

\subsection{Architecture} 
\begin{figure}
    \centering{\includegraphics[width=3in,height=1.0in]{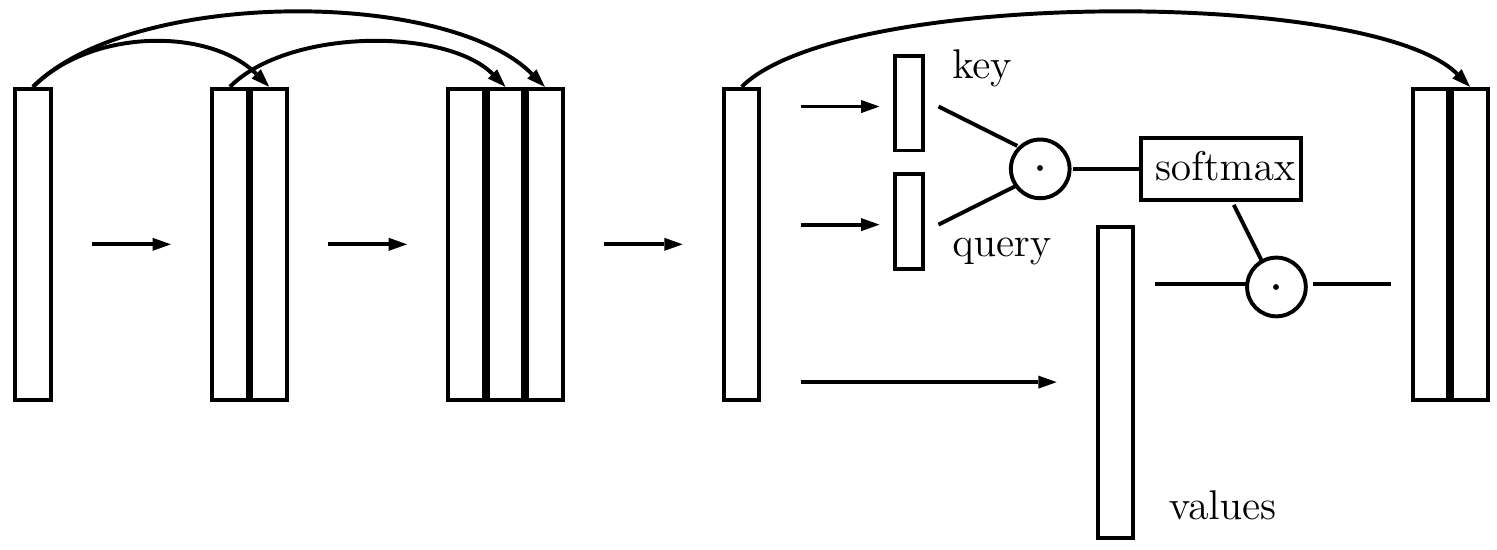}}\caption{Details of the DenseNet and attention block. Straight arrows are convolutions with batch normalization and ReLU non-linearity, curved arrows are concatenations. }
\label{figdenseattention}
\vspace{-0.15in}
\end{figure}

For the neural model $\mathcal{N}$, we start with a fully-convolutional 2D U-Net architecture with self-attention layers and 4-layer DenseNet blocks at each level, similar to \cite{tolooshams19}. We take the convolutions to be causal in the time direction, but not in the frequency direction, meaning that padding is applied symmetrically in the frequency direction as is usual in 2D convnets, but applied asymmetrically in the time direction in the sense that it is only used at the edge of each layer corresponding to the early part in time. This helps preserve the output quality at the late-portion of the output which is used in low-latency application as padding tends to hurt quality near edges and borders. Note that look-ahead is provided by the average-pooling layers, which are used instead of max-pooling. Figure \ref{figunetandpos} shows the overall architecture, while Figure \ref{figdenseattention} shows details of the DenseNet and attention blocks. 

The self-attention blocks we use are the same as the ones used in \cite{wang2018non, zhang2018self} with the exception that the mechanism aggregates information only in the time direction to increase efficiency during training and inference. 

\subsubsection{Frequency-Positional Embeddings} For early convolutional layers to be able to do frequency-aware processing, we concatenate a vector of positional embeddings to each time-frequency bin at the input layer of the model. The frequency-positional embedding vector for time-frequency bin centered at $(t, f)$ depends only on $f$ and is defined as:
$$\rho(t, f) = (\op{cos}(\pi \frac{f}{F}), \op{cos}(2\pi \frac{f}{F}), \dots, \op{cos}(2^{k-1}\pi \frac{f}{F})),$$
where $F$ is the frequency bandwidth and $k=10$.

\subsection{Datasets}

For the clean signal $s$, we combine data from two sources. First, we take two subsets of the publicly available LibriVox dataset, totaling approximately 600 hours of speech data: the LibriSpeech-clean\footnote{'clean' clips in LibriSpeech were selected based on WER in ASR experiments, not directly by audio quality}, \cite{librispeech}, as well as the subset of the LibriVox dataset filtered based on Mean Opinion Scores to form the DNS Challenge dataset \cite{dnschallenge}. The second source is VoxCeleb2 from which we use approximately 800 
hours of data.

\subsubsection{Semi-Supervised Learning with the VoxCeleb2 Dataset} To be able to use this large and varied dataset, we first train two models on the LibriSpeech dataset described above. The first model is a speech enhancement model that also does full dereverberation that is trained to estimate the reverb-only portion $h * s - s$, along with the clean signal $s$ and noise $n$. This model uses the same architecture as our proposed network, but we use fewer filters, and early stopping to avoid overfitting. We use this model to estimate the direct-to-reverberant ratio (DRR) of each clip in VoxCeleb2 and filter out clips with DRR less than 30 dB. While this model is better at estimating DRR compared to more traditional methods, 
its clean signal estimates contain artifacts and are not suitable for training. Instead, we use a second model with the same architecture, trained to estimate $h * s$ and $n$ only. We use this denoise-only model to filter out all clips with signal-to-noise ratio (SNR) less than 10 dB, and use its clean speech estimates as training data for subsequent experiments.

\subsubsection{Noise data and over-sampling nonstationary noise} For noise data, we filter the AudioSet dataset, selecting clips with tags from the AudioSet ontology that are sounds that a speech enhancement system would be expected to remove, while excluding any clips with tags related to sounds that humans make. 

We found that, even though most AudioSet tags correspond to non-stationary noise categories, a random 1-second chunk we use in training will more often than not have no non-stationary noise. We compute, for each chunk, the energy levels in 50 ms windows, and upsample, during training, chunks that have a standard-deviation of windowed energy of at least 3 dB. This increase the prevalence of non-stationary noise during training.

\subsection{Augmentations}

We use the following augmentation stack. Unless specified otherwise, distributions are uniform in the given number ranges. 
\begin{itemize}
    \item {\bf EQ}. Random high and low-shelf EQ filters. With center frequency chosen uniformly in logarithmic domain between 40 and 8000 Hz, gain between $\pm 10$ dB. Two random EQ bell-curves per datapoint, symmetric in log domain, with Q-value between 0.5 and 1.5; frequency chosen from the same interval as shelf EQ. Randomized and applied to both speech and noise separately. 
    \item {\bf Pitch shifts}. Random resampling with $\pm 10\%$ of the original sample rate.
    \item {\bf Clipping}. Random clipping between 0.5 and 1 of the peak value of the signal, applied 10\% of the time. 
    \item {\bf Empty buffer simulation}. Random deletion of the first 0.5 to 1 of the input signal to simulate partially filled buffer in low-latency evaluation. 
    \item {\bf Level and Silence}. We skip datapoints with foreground RMS less than -38 dBFS (dB relative to full-scale of $1.0$) and normalize each signal to have RMS value of -20 dBFS. We then apply a random volume down between -30 and 0 dB to the background, normalize the mix to -20 dBFS RMS, then apply a random amplification between -25 to 5 dB to everything. We additionally use silence as the foreground 3\% of the time.
    \item {\bf Band-limiting}. To make the model robust to cases where the input signal is band-limited, we apply a low pass filter at a frequency between 4 and 7 kHz, 2.5\% of the time to background only, 2.5\% of the time to foreground, and 5\% of the time to both. 
    \item {\bf Reverberation}. Used both as an augmentation and for datapoint creation as described below.  
\end{itemize}

\subsubsection{Reverberation} 

When adding reverberation, we first identify, in each  Room Impulse Response (RIR), the portion corresponding to the direct path, i.e. the 'first tap', and scale and shift the RIR so that the first tap is at $t=0$ and it has height 1. So we have $x = s * (h_{0} + h_{>0}) + n$ where $h_0$ is a single tap at time zero. We then apply a gain to all taps except the first tap by a value between -25 and 0 dB. Also, 60\% of the time, we add reverberation via the same impulse response to the noise signal as well, except that there is a separate downward scaling of the non-first tap. Hence, the model input becomes
$$ x = s * ({{h}_{0}} + \alpha h_{>0}) + (n * (h_0 + \beta h_{>0})).$$

We use both real-recorded and synthetic room impulse responses (RIRs). For real impulse responses, we use the Aachen Impulse Response dataset \cite{jeub2009binaural} consisting of of 214 RIR recordings. For synthetic RIRs, we generate a library of 10,000 RIRs, using the image method \cite{allen1979image}, with random rectangular rooms with sizes from 2 to 10 meters with random reflection coefficients between $0.5$ and $1.5$.

We restrict to using impulse responses with $\op{RT}_{60} < 0.8\op{s}$. We further augment all impulse responses with random resampling, which simulates changing room sizes with the same materials, and random exponential decays, which approximate changing uniform absorption levels of the room material. 

\subsubsection{Partial or No Dereverberation} We experiment with no-dereverberation models, where, during training, reverberation is used simply as an augmentation, and the foreground speech label is $y = s * h$; and with partial-dereverberation, where the label's room impulse response has the first $20$ms unaltered, and then made to decay quickly, to make $\op{RT}_{60} < 0.2\op{s}$, by multiplying with an exponential decay function.

\subsection{Loss functions}

We train the neural model by optimizing, for each target $y$, the loss function
$$\cL(y, \widehat{y}) = \lambda_{\text{audio}}\cL_{\text{audio}}(y, \widehat{y}) + \lambda_{\text{spectral}}\cL_{\text{spectral}}(Y, \widehat{Y}),$$
where the audio loss is the L1 loss,
$$ \cL_{\text{audio}}(y, \widehat{y}) = |y - \widehat{y}|.$$ 
For the spectral loss function $\cL_{\text{spectral}}$, let $Y_{t, f} = |\text{STFT}(y)_{t, f}|$ and $\widehat{Y}_{t, f} = |\text{STFT}(\widehat{y})_{t, f}|$, be the $\op{STFT}$ bin magnitudes; we set
\begin{equation*}
\begin{aligned}
 & \cL_{\text{spectral}}  = \\ 
 & \sum_{t, f}\op{w}(f)\left(\lambda_{\text{over}}\mathbb{1}_{\widehat{Y} \geq Y, t, f} + \lambda_{\text{under}}\mathbb{1}_{\widehat{Y} < Y, t, f} \right)|Y_{t, f} - \widehat{Y}_{t, f}|. 
\end{aligned}
\end{equation*}
Here, $\op{w}$ is a frequency-weighting function, and $\mathbb{1}_{\widehat{Y} \geq Y, t, f}$ is the characteristic function with value 1 if $\widehat{Y}_{t, f} \geq Y_{t, f}$ and value 0 otherwise. The variables $\lambda_{\text{over}}$ and $\lambda_{\text{under}}$ bias the model for overestimation or underestimation of the speech magnitude.

\subsection{Inference-Time} 

\begin{figure}[htb!]
    \centering{\includegraphics[width=2.8in]{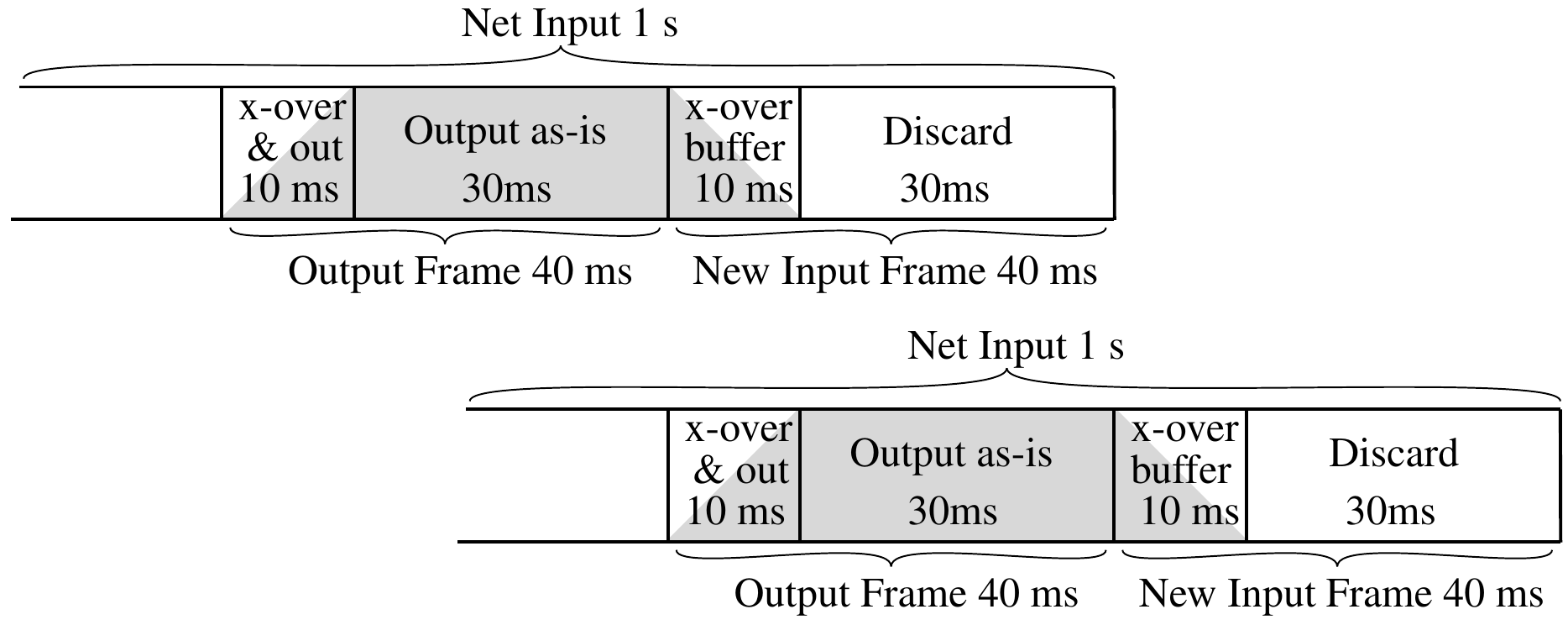}}\caption{The Inference-time mechanism. The convnet is evaluated always on the last second of the input buffer. }
\label{figinference}
\end{figure}

\begin{table*}[hbt!]
\footnotesize
\centering
\begin{tabular}{lcccc} 
\hline \\[-2ex]
$\,$& \bf{Full} & Synthetic no Reverb & Synthetic with Reverb & Real Recordings \\
\hline \\[-2ex]

Proposed & \bf{3.833} & \bf{4.160} & \bf{3.613} & 3.779 \\
\hspace{0.07in} without dereverberation & 3.807 & 4.128 & 3.497 & \bf{3.802} \\
\hspace{0.07in}  without positional embeddings & 3.789 & 4.092 & 3.548 & 3.758 \\
\hspace{0.07in}  without background reverbs & 3.768 & 4.119 & 3.573 & 3.690 \\
\hspace{0.07in}  without biased loss & 3.766 & 4.094 & 3.519 & 3.726 \\
\hspace{0.07in}  without semi-supervised data & 3.755 & 4.131 & 3.620 & 3.634 \\
\hspace{0.07in}  without reverb augmentations & 3.467 & 4.133 & 2.358 & 3.688 \\
RNNoise \cite{valin2018rnnoise} & 3.464 & 3.660 & 3.162 & 3.517 \\
DNS Challenge Baseline \cite{dnschallenge} & 3.439 & 3.703 & 3.120 & 3.466 \\
Noisy & 3.432 & 3.568 & 3.183 & 3.489\\
\hline 
$95\%$ confidence interval & $\pm 0.04$ & $\pm 0.06$ & $\pm 0.08$ & $\pm 0.05$ \\
\hline 
\end{tabular}
\vspace{0.05in}
\caption{Mean Opinion Score evaluation of different Algorithms over the DNS Challenge non-blind test sets. } 
\label{tablemos}
\end{table*}

\begin{table*}[hbt!]
\vspace{-0.1in}
\footnotesize
\centering
\begin{tabular}{l|cccc|cccc} 
\hline \\[-2ex]
$\,$ & \multicolumn{4}{c|}{Synthetic without Reverb} & \multicolumn{4}{c}{Synthetic with Reverb}\\
\hline \\[-2ex]
Methods  &{PESQ} &{CBAK} &{COVL} &{CSIG}  &{PESQ} &{CSIG}  &{CBAK} &{COVL} \\
\hline

RNNoise \cite{valin2018rnnoise} 
& 1.973 & 3.463 & 2.789 & 2.692 & 1.777 & \bf{3.407} & 2.709 & 2.569 \\
DNS Challenge Baseline \cite{dnschallenge}    &
1.811 & 2.003 & 2.235 & 2.781 &
1.515 & 1.937 & 1.949 & 2.515 \\
Noisy & 1.582 & 2.533 & 2.350 & 3.186 & 1.821 & 2.801 & 2.635 & 3.499 \\
\bf{Proposed} & 2.745 & 3.040 & \bf{3.422} & \bf{4.080} & 1.609 & 2.303 & 2.223 & 2.906 \\
\hspace{0.07in}  without reverb augmentation & \bf{2.748} & \bf{3.043} & 3.366 & 3.965 & 1.293& 1.935 & 1.582 & 2.017 \\
\hspace{0.07in}  without semi-supervised data	& 2.722 & 3.022 & 3.343 & 3.94 & 1.612 & 2.302 & 2.214 & 2.887 \\
\hspace{0.07in}  without positional embeddings	& 2.721 & 3.012 & 3.358 & 3.982 & 1.638 & 2.312 & 2.188 & 2.807 \\
\hspace{0.07in}  without partial dereverberation	& 2.707 & 3.015 & 3.286 & 3.851  & \bf{2.832} & 3.209 & \bf{3.349} & \bf{3.834} \\
\hspace{0.07in}  without background reverbs	& 2.667 & 2.997 & 3.296 & 3.909 & 1.613 & 2.301 & 2.223 & 2.906 \\
\hspace{0.07in}  without biased loss & 2.457 & 2.904 & 3.106 & 3.749 & 1.545 & 2.258 & 2.070 & 2.671 \\
\hline 
\end{tabular}
\vspace{0.05in}
\caption{Objective evaluation of different Algorithms over the DNS Challenge synthetic non-blind test sets. Note that the Synthetic with Reverb test reference clean labels contain reverberation, results the model that is trained to keep all reverberation has the best performance on this set. 
}
\label{tableobjectivemetrics}
\vspace{-0.25in}
\end{table*}

For low-latency evaluation, we use 40 ms-sized input frames (i.e. 640 samples at 16kHz) with one-frame look-ahead. For each input chunk of samples, we run the model on the last 16384 samples in the input buffer. We use cross-over to eliminate artifacts from the frames. Figure \ref{figinference} illustrates the evaluation mechanism for two input chunks. Inference takes 0.65 seconds per second of audio on a V100 GPU.


\section{Experiments and Results}

For each model, we use 6 down-blocks in the U-Net with the number of per-layer filters in each being 32, 64, 128, 256, 256, 256, and up-blocks symmetric to the down-blocks. There are a total of $\sim$50M parameters. We multiply the foreground and background losses with weighting coefficients $\lambda_{\text{fg}} = 2.0$, and $\lambda_{\text{bg}} = 0.4$ for the foreground and background estimation respectively. We take $\lambda_{\text{audio}} = 1$, and $\lambda_{\text{spectral}} = 1.5$, with (only for the foreground signal) $\lambda_{\text{over}} = 2.6$ and $\lambda_{\text{under}} = 13.3$. We have not used any hyper-parameter tuning techniques, with most parameters, especially those used for augmentations, set as sensible defaults and unmodified. 

We train each model for 700,000 iterations with total mini-batch size of 112, using the ADAM optimizer, with learning rate of 1e-4, halved every 100,000 iterations. Training each model takes about 4 days on 8 V100 GPUs. Our implementation uses the MXNet \cite{mxnet} framework. 

\subsection{Evaluation: Subjective and Objective Metrics}

For human opinion tests, we use the methodology of ITU-T P.808 \emph{Subjective Evaluation of Speech Quality with a Crowdsourcing Approach} \cite{P.808}, on Amazon Mechanical Turk.The MOS scores are based on 10 listenings each of the model's outputs on the 600 real and synthetic inputs on the INTERSPEECH 2020 DNS Challenge test set \cite{dnschallengefinal}, which covers varied cases.  

For objective metrics, we evaluate wide-band Perceptual Evaluation of Speech Quality (PESQ) -- ITU-T P.862.2 -- \cite{P.862}, and the composite CSIG, CBAK, and COVL scores proposed in \cite{loizou2007speech} on the 300 synthetic examples in the same test set.


\subsection{Results}

Tables \ref{tablemos} and \ref{tableobjectivemetrics} show, respectively, objective and subjective metric evaluation results. Note that, removing the semi-supervised conversational data has a strong effect on performance on real recordings which tend to have more varied speech styles. 

The proposed model took 1st place in the 2020 Deep Noise Suppression Challenge's Non-Real-Time Track \cite{dnschallengefinal}. 

\begin{table}[hbt!]
\footnotesize
\centering
\begin{tabular}{lcccc} 
\hline \\[-2ex]
$\,$& \textbf{Full} & Synthetic & Synthetic & Real\\
$\,$& & without Reverb & with Reverb & Recs \\

\hline \\[-2ex]
Noisy & 2.95 & 3.13 & 2.64 & 2.83 \\
\bf{Proposed}  & 3.52 & 4.07 & 3.19 & 3.40 \\
\hline 
\end{tabular}
\vspace{0.05in}
\caption{Mean Opinion Score evaluation provided by the DNS Challenge based on the blind test set. Full results are available in \cite{dnschallengefinal} (team \#9). } 
\label{tablechallengemos}
\vspace{-0.3in}
\end{table}

\section{Conclusion}

We described new techniques that result in improvements to speech enhancement with large neural networks. The resulting PoCoNet speech enhancer is a large U-Net with DenseNet and self-attention blocks with frequency-positional embeddings, and is trained with a semi-supervised technique partly on conversational data, with an extensive augmentation stack including reverberation, and with a loss function that is biased to preserve speech. Evaluation results show the quality improvement on the overall system due to each component and demonstrate the effectiveness of the introduced techniques for training large neural speech enhancers.

\bibliographystyle{unsrt}
\bibliography{bibliography}

\end{document}